# BCS-type second-order phase transition of classical Langmuir wave system

Eiichirou Kawamori

Institute of Space and Plasma Sciences, National Cheng Kung University, Tainan, Taiwan

**ABSTRACT**

The BCS-type second-order phase transition of a classical Langmuir wave system is identified theoretically and numerically. The transition takes place between two states: Langmuir wave turbulence (LWT) and Langmuir wave supercontinuum (LWSC), the latter of which exhibits broad power spectra with significant spatiotemporal coherence when a certain number of plasmons (plasma wave quanta) are excited in the system. In the LWT-LWSC transition, the modulational instability and resulting creation of plasmon pairs are the classical counterparts of the Cooper instability and creation of Cooper pairs in superconducting states. The Bose–Einstein condensation of Cooper pairs in superconducting states is replaced by the Kuramoto oscillator-entrainment of plasmon pairs in a LWSC. The order parameter of the LWSC state, which is defined as the mean field of the plasmon pairs, takes on a significant value, which clearly indicates that a macroscopic number of plasmon pairs occupy a single momentum state with an identical phase in the LWSC. The emergence of spatiotemporal coherence and the decrease in the phase randomization are considered as development of long-range order and spontaneous symmetry breaking, respectively, indicating that the LWT-LWSC transition is a second order phase transition phenomenon. By this transition, U(1) symmetry of LWT is broken. The LWSC is, in terms of plasma physics, a so-called Bernstein–Greene–Kruskal mode characterized by its undamped nature.

The nonlinear Schrödinger equation (NLSE) is an integrable nonlinear partial differential equation that describes waves in nonlinear media. It is applicable to various physical systems regardless of their detailed properties. Phenomena described by the NLSE are exemplified by gravity waves in neutral fluids, Langmuir waves (LWs) in collisionless plasmas [1], and solitons in Bose–Einstein condensates [2]. For initial-value problems, the NLSE may be solved by applying the inverse scattering method [3]. Because it can be integrated, the NLSE may be described as a Hamiltonian system. However, in real physical systems, mechanisms exist that prevent integration, such as dissipation stemming from interaction with the environment and higher-order dispersion. Specifically, although LWs with finite amplitudes have been described by the integrable NLSE in a fluid picture, LW systems should actually be regarded as an open system of plasmons (LW quanta) in environments consisting of particles (ions and electrons). Therefore, unlike conventional NLSE systems, the total number of plasmons is indefinite [4], and such systems should be described by a grand potential rather than by a Hamiltonian. In this article, I show theoretically and numerically that one-dimensional (1D) LW systems described by the NLSE incorporating a grand potential exhibit a second-order phase transition between a Langmuir wave turbulence (LWT) and a coherent Langmuir wave supercontinuum (LWSC) state [5]. This LWT-LWSC transition is well described by the proposed theory, which is a classical version of the BCS theory that describes the superconducting transition [6]. The present theory provides a new insight into systems conventionally described by the NLSE, including rogue waves in neutral fluids [7] and optical turbulence [8]. Indication of this kind of transition can be seen, in addition to the LWT-LWSC transition, in the capillary rogue waves observed in a laboratory experiment [7]. The maximum tricoherence of the surface elevation of water showed a discontinuous transition from the weak phase coupling in four-wave interactions to the strong one when the forcing of the water container exceeded the threshold [7].

I develop the theory of the LWT-LWSC transition based on observations of a 1D particle-in-cell (PIC) simulation of LWs. To begin, I report some remarkable observations of LWSC generation and of the LWT-LWSC transition in the 1D PIC simulation.

The PIC code is spatially 1D and follows the particle velocities in 1D. The parameters of the simulation are as follows: the time step $\Delta t$, grid size $\Delta x$, number $\Delta N$ of particles (ions and electrons) per grid cell, and ratio $m_i/m_e$ between ion and electron mass are $0.03\omega_{pe}^{-1}$, $0.3\lambda_{de} = 0.3 v_{the}\omega_{pe}^{-1}$, $2 \times 1 \times 10^4$, and $1836.0$, respectively, where $\omega_{pe}$, $\lambda_{de}$, and $v_{the}$ are the plasma angular frequency, Debye length, and thermal velocity of the electrons, respectively.

I apply an external drive field as a pump (seed) in the initial phase ($t = 0$–$900\ \omega_{pe}^{-1}$) of the simulations by adding a drive term $E_{drive}$ to the self-consistent electric-field term in the equation of motion of the electrons and ions. To generate LWT and LWSC states, a sinusoidal driver with frequency $\omega$ varied as $\omega/\omega_{pe} = 1.10, 1.24$, and $1.30$ with a fixed wavenumber $k_{drive} = 2\pi/5\ [\lambda_{de}^{-1}]$ is applied to Maxwellian plasmas with periodic boundary conditions and with the computational domain $x = 0\lambda_{de}$–$600\lambda_{de}$. For $k_{drive}$, a frequency of the linear modes of the target plasma (the least-damped Landau root) is $\omega/\omega_{pe} = 1.24$.

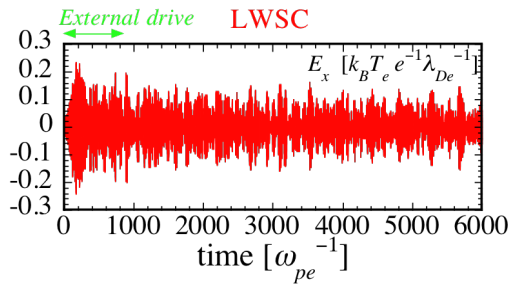

Fig. 1 Time evolution of longitudinal electric field $E_x$ measured at $x = 0.3\ \lambda_{de}$ for a LWSC state. From $t \sim 2700\ \omega_{pe}^{-1}$, a pulse train with an interval of approximately $400\ \omega_{pe}^{-1}$ is observed, indicating spectral broadening of $E_x$ with phase locking.

Figure 1 shows time evolution of longitudinal electric field $E_x$ measured at $x = 0.3\ \lambda_{de}$ for a LWSC state. Although it seems to be irregular waveform, a pulse train with an interval of approximately 400 $\omega_{pe}^{-1}$ is observed from $t \sim 2700\ \omega_{pe}^{-1}$, indicating spectral broadening of $E_x$ with phase lock.

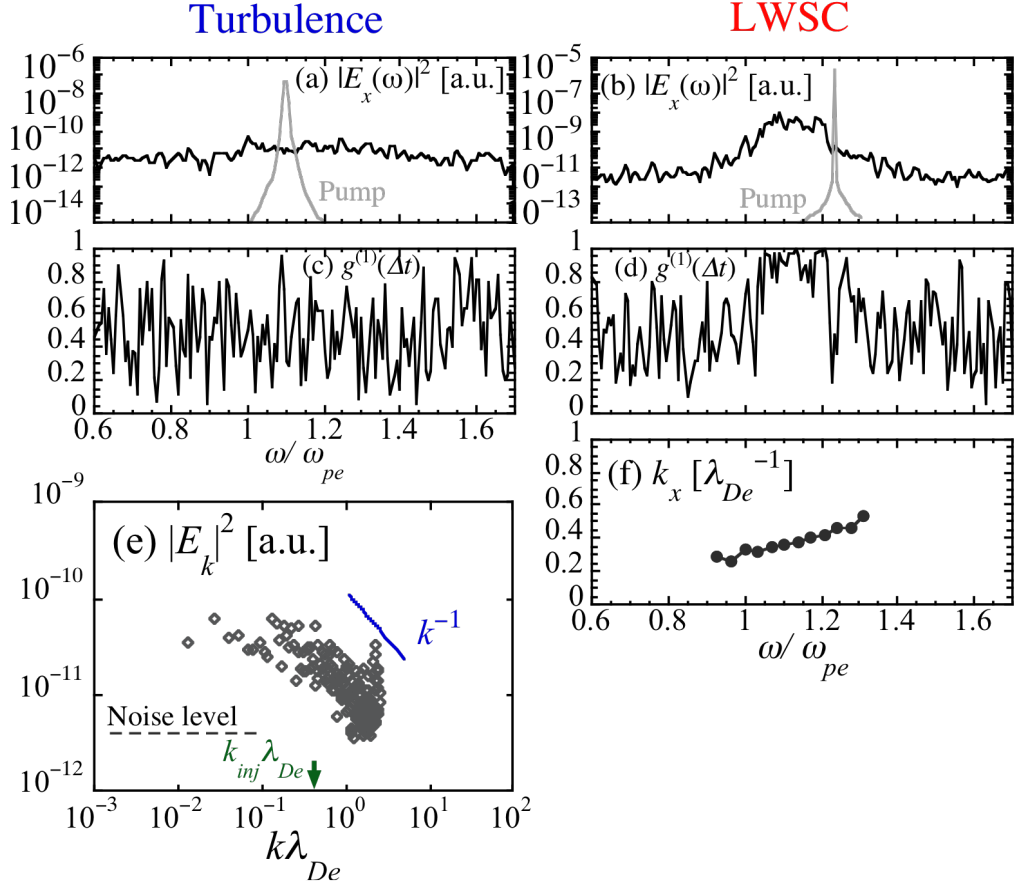

Fig. 2. (a, b) Power spectra and (c, d) first-order coherence $g^{(1)}$, for LWT and LWSC, respectively. (e) Wavenumber spectrum of LWT. (f) Wavenumber as a function of $\omega/\omega_{pe}$ for LWSC. All graphs are calculated for a longitudinal electric field $E_x$.

Figures 2(a), 2(b) and 2(c), 2(d) show the power spectra $|E_x(\omega)|^2$, including that of pump waves (gray) and the first-order coherence $g^{(1)}$, respectively, for the LWT and LWSC. Figure 2(e) shows the wavenumber spectrum of the LWT, and Fig. 2(f) shows the wavenumber as a function of $\omega/\omega_{pe}$ for the LWSC. All graphs in Fig. 2 are calculated for a longitudinal electric field $E_x$. The Hanning window is applied to the Fourier analyses to suppress spurious coherence with side lobes. The first-order coherence $g^{(1)}$ is a measure of temporal coherence of the wave field, whose range is $0 \leqq g^{(1)} \leqq 1$, and is defined as

$$\left|g^{(1)}(\omega, t_2 - t_1)\right| = \left|\frac{\langle E_x^*(\omega, t_1) E_x(\omega, t_2)\rangle}{\sqrt{\langle |E_x(\omega, t_1)|^2\rangle}\sqrt{\langle |E_x(\omega, t_2)|^2\rangle}}\right|.$$

$E_x(\omega, t)$ and $\omega$ are the Fourier component of $E_x$ in the time window $t$ ($t_1 = 2520$, $t_2 = 3420$ $\omega_{ee}^{-1}$) and the angular frequency of the Fourier modes, respectively. In the simulations, $g^{(1)}$ is regarded as an indicator of spatiotemporal coherence of the waves because the simulation adopts periodic boundary conditions. $g^{(1)}$ is unity when the wave at the frequency component completely maintains phase coherence during the specified duration. The LWSC has a broadened power spectrum with a peak whose frequency is downshifted from the pump frequency, $\omega = \omega_{\text{carrier}} \equiv 1.24\, \omega_{pe}$, and the significantly high $g^{(1)}$ in the spectral range of 1.0–1.20$\omega_{pe}$, contrary to the incoherent spectrum of the LWT, as shown in Figs. 2(a) and 2(c). Figure 2(e) shows normal and inverse cascades from the wavenumber $k_{inj}$ of the external drive, exhibiting a typical turbulent nature. The solid lines in Fig. 2(e) represent Kolmogorov–Zakharov scaling with artificial substitution of $d = 1$ to the spatial dimension in the scaling law under the condition that constant energy input exists at a fixed $k$, despite the fact that the Kolmogorov–Zakharov theory is inapplicable to $d = 1$ [8–10]. As shown in Fig. 2(f), for the LWSC, $\omega$ can be approximated as $\propto k$ within the SC frequency band. This fact is crucial for developing the theory.

Figure 3(a) shows the full width of the SC frequency band $w_{coh}$, which is defined as the bandwidth in which $g^{(1)} > 0.5$. Figures 3(b) and 3(c) show, respectively, the classical von Neumann entropy $S_{\text{Neumann}}$ of $E_x(t)$ and the order parameter $|m_{\text{Kuramoto}}|$ as functions of the total wave quanta of the plasmons, $N_{\text{plasmon}} \equiv \int |E_x(\omega)|^2 d\omega$. $S_{\text{Neumann}}$ is a measure of phase-randomizing degree of $E_x(t)$ [5], which is defined as $S_{\text{Neumann}} \equiv -\text{Tr}(\rho \ln \rho)$, where Tr represents the trace of the matrix and $\rho$ is the density matrix of $E_x(t)$, which is defined as $\rho \equiv \sum_k p_k \left\{ \sum_i^{N\text{mode}} \sum_j^{N\text{mode}} a_i a_j^* |\omega_i\rangle\langle\omega_j| \right\}_k$, where $a_i$ is the Fourier coefficient of mode $|\omega_i\rangle$ and $p_k$ is the provability of realizing the state $k$ [5]. $|m_{\text{Kuramoto}}|$ is a fraction of wavenumber-condensed plasmon pairs with identical phase, ranging from 0 to 1 (its strict definition is given by Eq. (20)). The various symbols (solid squares, gray circles, and open circles) represent cases for angular frequencies of external drive fields. The LW system exhibits a clear transition from turbulent states with zero $w_{coh}$ to coherent SC states having finite values of $w_{coh}$ accompanied by a decrease in $S_{\text{Neumann}}$ as $N_{\text{plasmon}}$ increases. The emergence of the long-range coherence shown by a finite value of $w_{coh}$ can be taken as the development of long-range order. The suppression of $S_{\text{Neumann}}$ reflects the fact that random phase states disappear; that is, the symmetry of the wave phase breaks, which is indicative of spontaneous symmetry breaking. This development of ordered structure is exactly captured by the relationship between $|m_{\text{Kuramoto}}|$ and $N_{\text{plasmon}}$. In the proposed theory, $N_{\text{plasmon}}$ is proportional to the square root of the interaction strength between the plasmon pairs, as will be shown below in the theory section. The enhancement of $|m_{\text{Kuramoto}}|$ indicates that a macroscopic number of plasmon pairs condense in a single state (i.e., single wavenumber and identical phase), which is a typical feature of superconductor phase transitions. Insets of Fig. 3(c) depict states of plasmon pairs in LWT and LWSC, respectively. The arrows represent wave electric fields of the plasmon pairs with those phases shown by the orientations of the arrows. In the LWSC states, a macroscopic number of plasmon pairs aligns in an identical direction whereas those are directed at random directions in the LWT states. In the LW system, the LWT and LWSC states correspond to the normal and superconducting states in the superconducting transition, respectively. A discussion of these features is provided below in the theory section.

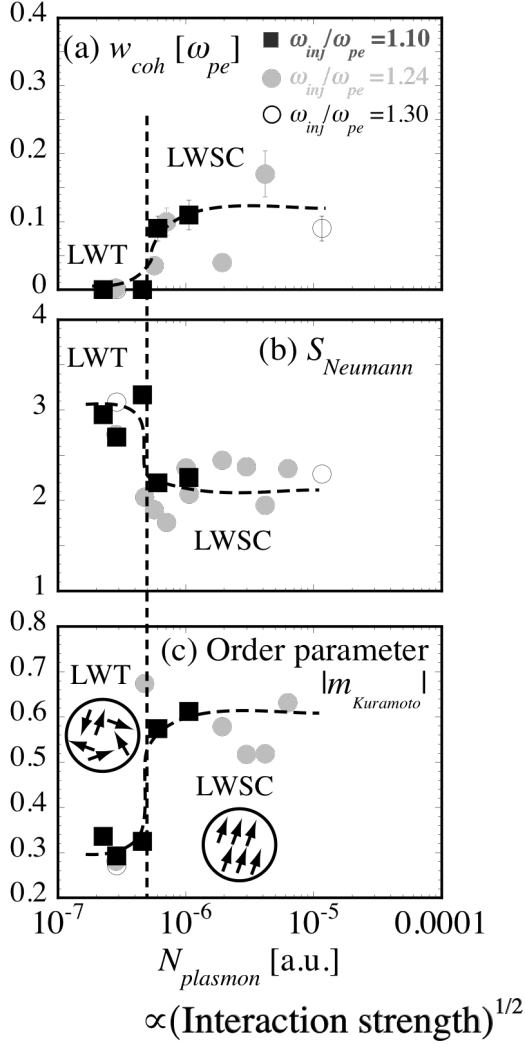

Fig. 3. (a) Full width $w_{coh}$ of the first-order coherence band having $g^{(1)} > 0.5$, (b) von Neumann entropy $S_{Neumann}$, and (c) order parameter $|m_{Kuramoto}|$ as functions of the total wave quanta of plasmons, $N_{plasmon}$. The various symbols represent the respective angular frequencies of external drive fields.

I now start developing a theory to illustrate the observed transition properties. As mentioned above, the integrable NLSE has been used to describe the spatiotemporal behavior of the envelope $a$ of LWs [1, 4]. Specifically, the NLSE takes the form

$$i\frac{\partial a}{\partial \tau} + p\frac{\partial^2 a}{\partial \varsigma^2} + q|a|^2 a = 0, \qquad (1)$$

where $\tau$ and $\zeta$ are the extended time spatial coordinate in a moving frame, respectively. Furthermore, $p \equiv (\partial \omega^2/\partial k^2)_0$ and $q \equiv (\partial \omega/\partial |a|^2)_0$, where $(\cdot)_0$ indicates the evaluation of the argument at the carrier wavenumber and amplitude. Equation (1) exactly describes integrable properties of LWs such as envelope solitons when the system can be regarded as an isolated system. In wavenumber space ($k$ space), Eq. (1) becomes [11, 12]

$$\frac{\partial \hat{a}_k}{\partial \tau} = -i\omega_k \hat{a}_k - iV \int \hat{a}_2^* \hat{a}_3 \hat{a}_4 \delta_{k+2-3-4}\, dk_2 dk_3 dk_4, \qquad (2)$$

where $V$ and $\delta_{k+2-3-4} \equiv \delta(k + k_2 - k_3 - k_4)$ are the interaction matrix and Dirac delta function, respectively. Here I assume $V = $ const. $\delta_{k+2-3-4}$ enforces the wavenumber-matching condition for the four-plasmon interaction $k + k_2 = k_3 + k_4$.

System (1) or (2) is a Hamiltonian system with the Hamiltonian [11, 12]

$$H = \int \omega_k |\hat{a}_k|^2 \, dk - V \int \hat{a}_1 \hat{a}_2 \hat{a}_3^* \hat{a}_4^* \delta_{1+2-3-4} \, dk_1 dk_2 dk_3 dk_4, \tag{3}$$

where $\omega_k$ and $\hat{a}_k$ are the linear dispersion frequency measured with respect to the carrier frequency $\omega_{\text{carrier}}$ (i.e., $\omega_k = \omega(k) - \omega_{\text{carrier}}$) and the Fourier modes of $a$, respectively. The Hamiltonian form of Eq. (1) is $i \frac{\partial \hat{a}_k}{\partial \tau} = \frac{\delta H}{\delta \hat{a}_k^*}$. The first term on the right-hand side (RHS) of Eq. (3) is the linear-mode energy of the plasmons, and the second term represents the interaction energy between the four plasmons. The Hamiltonian (3) is one of the integrals of the NLSE (1). In addition to the Hamiltonian, the total number of plasmons, $N_{\text{plasmon}} = \int |\hat{a}_k|^2 \, dk$, is also conserved. However, actual LWs are physical systems in which waves and particles coexist, so the total number of plasmons is indefinite, unlike the integrable NLSE. To handle this problem, I phenomenologically introduce a chemical potential $\mu = \mu_F + \omega_{\text{carrier}}$ by adding a term $-\mu_F N_{\text{plasmon}} = -\mu_F \int |\hat{a}_k|^2 \, dk$ to the RHS of Eq. (3), resulting in the replacement $\omega_k \to \xi_k \equiv \omega_k - \mu_F$. $\mu_F$ is the chemical potential measured from $\omega_{\text{carrier}}$ and a function of $N_{\text{plasmon}}$ and is considered to be the frequency at which plasmons are created, in other words, the (nonlinearly-shifted) least-damped Landau root. The product $\hbar \mu$ corresponds to the Fermi energy in conventional BCS superconductor theory. For simplicity, $\mu_F$ is hereafter treated as a fixed value. With this prescription, $H$ becomes the thermodynamic potential at zero temperature.

$$H = \int \xi_k |\hat{a}_k|^2 \, dk - V \int \hat{a}_1 \hat{a}_2 \hat{a}_3^* \hat{a}_4^* \delta_{1+2-3-4} \, dk_1 dk_2 dk_3 dk_4. \tag{4}$$

Although the chemical potential $\mu_F$ is introduced phenomenologically in this study, it should be deduced by using kinetic theory in a consistent manner.

If $V \hat{a}_1 \hat{a}_2 \hat{a}_3^* \hat{a}_4^*$ is positive (i.e., $q/p > 0$), the interaction between four plasmons is attractive. In the superconductor theory, this attractive interaction induces the Cooper instability, in which a pair of electrons having a zero total momentum at the Fermi surface occupy a state whose energy is lower than twice the Fermi energy [13]. For the LW system, the Cooper instability is replaced by a modulational instability (MI) [14]. The unstable condition for the MI is $pq > 0$ [14]. Owing to the MI, two plasmons created at frequency $\mu_F$ are preferentially redistributed (scattered) into the neighboring frequencies having zero total momentum (relative to the Fermi wavenumber $k_f$ corresponding to $\omega_k = \mu_F$):

$$2k_F = (k_F + \Delta k) + (k_F - \Delta k), \tag{5}$$
$$2\mu_F = \xi_{k_F + \Delta k} + \xi_{k_F - \Delta k}. \tag{6}$$

This means that when nonlinearity becomes significant in LW systems, a state exists that is lower in energy than the linear states represented by the first term in Eq. (4). Namely, the system prefers to redistribute plasmons with the linear-mode frequency into neighboring frequencies whose average is equal to $\mu_F$. Note that $\omega_k \propto k$ in the LWSC system, as shown in Fig. 2(f). Therefore, consideration of the behavior of the LW systems in $k$ space can be replaced with that in $\omega$ space. The frequency band with zero width at the frequency $\mu_F$ corresponds to the Fermi sphere in the superconductor theory, although plasmons in our system are considered classical quasi-particles.

I make two assumptions about plasmon-pair scattering. (i) The center-of-mass momenta (wavenumbers) of plasmon pairs involved in four-wave mixing (FWM), which is represented by the second term on the RHS of Eq. (4), are zero in the $k_f$ frame. That is, plasmon pairs whose total wavenumbers $k_{\text{total}} = k_1 + k_2 = 2k_F$ are scattered into new plasmon pairs having $k_{\text{total}} = k_3 + k_4 = 2k_F$. Thus, I retain only the components $\frac{(k_1+k_2)}{2} = \frac{(k_3+k_4)}{2} = k_F$ in the following mean-field approximation. (ii) The coefficient $V$ of the attractive interaction of MIs is finite and constant within the range $\mu_F - \omega_D \leq \omega_k \leq \mu_F + \omega_D$ and zero outside of this range. In the superconductor BCS theory, $\omega_D$ is the frequency of a lattice oscillation (i.e., a phonon). Therefore, superconductor transition temperature depends on the ion mass of the background lattice, which is called the isotope effect. However, in the LW system, $\omega_D$ may be determined as the upper bound (or the lower bound) of the instability band of the MI. According to the results of our PIC simulation, $\omega_D$ is independent of ion mass.

To verify assumptions (i) and (ii), I apply a tricoherence analysis [15] to the LWSC in the PIC simulation. The tricoherence $|t|^2$ is a measure of the fraction of the total product of powers of the frequency quartet ($\omega_1, \omega_2, \omega_3, \omega_1+\omega_2-\omega_3$) that is caused by cubically phase-coupled modes and is given by $|t|^2(\omega_1, \omega_2, \omega_3) = |T(\omega_1, \omega_2, -\omega_3)|^2$, where $T$ is the trispectrum and is defined as

$$|T(\omega_1, \omega_2, -\omega_3)| = \frac{E[X(\omega_1)X(\omega_2)X(-\omega_3)X^*(\omega_1+\omega_2-\omega_3)]}{\sqrt{P_{1,2}(\omega_1,\omega_2,-\omega_3)P(\omega_1+\omega_2-\omega_3)}}.$$

In this expression, $E$ is the expectation operator that averages over an ensemble of realizations, $X$ is the Fourier transform of a realization of the time series data, and $P$ is the power and is defined as $P_{1,2,3}(\omega_1, \omega_2, -\omega_3) = E[X(\omega_1)X(\omega_2)X(-\omega_3)X^*(\omega_1)X^*(\omega_2)X^*(-\omega_3)]$. Figure 4 shows $|t|^2$ of $E_x$ for the LWSC state shown in Fig. 1, with $\omega_3/\omega_{pe}$ chosen as the peak frequencies $\omega_{\text{peak}}$ of the power spectrum (i.e., $\omega = \mu = 1.15\omega_{pe}$). For this calculation, I used 30,000 samples as the length of a realization (which results in a frequency resolution of $\Delta\omega = 2\pi\omega_{pe0}/900$) and a realization number of 20. This plot represents the coupling between four waves that satisfy the matching conditions $\omega_1 + \omega_2 = \omega_3 + \omega_4$ and $k_1 + k_2 = k_3 + k_4$ ($\because \omega_k \propto k$). Note that the horizontal and vertical stripes at $\omega_1/\omega_{pe0} = \omega_3/\omega_{pe0}$ and $\omega_2/\omega_{pe0} = \omega_3/\omega_{pe0}$ showing high $|t|^2$ in Fig. 4 have no physical meaning because $|t|^2$ becomes always unity when those values are substituted in the definition. The significantly high $|t|^2$ region follows the line $\omega_1 + \omega_2 = 2\omega_3 = 2\mu_F$ (i.e., $(k_1 + k_2)/2 = (k_3 + k_4)/2 = k_F$) in Fig. 4. In addition, one can know that the FWM region is confined in the vicinity of $\omega_3 = 2\mu_F$ with a finite bandwidth, which agrees with our assumption of a finite interaction band $\mu_F - \omega_D \leq \omega_k \leq \mu_F + \omega_D$, where $\mu - \omega_D \sim 1.08\omega_{pe}$ and $\mu + \omega_D \sim 1.24\omega_{pe}$, respectively, in this example.

By following the above observation associated with our hypothesis, it is natural to introduce the following mean-field approximation to the plasmon pairs $\hat{a}_j\hat{a}_{-j}$ in the Hamiltonian (4):

$$\Delta = |\Delta|e^{i\theta} \equiv V\sum_j\langle \hat{a}_{-j}\hat{a}_j\rangle, \tag{7}$$

$$\Delta^* = |\Delta|e^{-i\theta} \equiv V\sum_j\langle \hat{a}_j^*\hat{a}_{-j}^*\rangle. \tag{8}$$

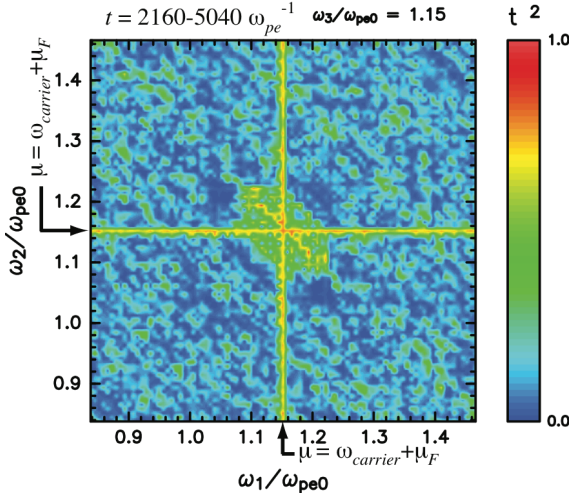

Fig. 4 Tricoherence $|t|^2$ of $E_x$ for LWSC shown in Fig. 1, with $\omega_3/\omega_{pe}$ chosen as the peak frequency $\omega_{peak} = \mu = 1.15\omega_{pe}$ of the power spectrum, where $\mu = \omega_{carrier} + \mu_F$. This result clearly confirms that scatterings following $\omega_1 + \omega_2 = 2\mu_F$ dominate.

Here the summation with respect to $j$ (wavenumbers) is implemented within the range $\mu_F - \omega_D \leq \omega_k \leq \mu_F + \omega_D$, and the brackets $\langle\rangle$ are the expectation value or ensemble average. The complex value $\Delta$ has two physical meanings: the order parameter and the energy (gap) necessary to dissociate plasmon pairs into two independent plasmons. The former meaning is understood straightforwardly from the definition given by Eq. (7); that is, a significant value for $\Delta$ indicates that a significant number of plasmon pairs have identical phase, which in turn indicates the condensation of a macroscopic number of plasmon pairs. The latter meaning is explained in detail two paragraphs below.

By using Eqs. (7) and (8) in Eq. (4) after applying the mean-field approximation, the Hamiltonian can be expressed as

$$H_{mf} = \sum_{k\geq 0}(\xi_k|\hat{a}_k|^2 + \xi_{-k}|\hat{a}_{-k}|^2) - \sum_{k\geq 0}(\Delta^*\hat{a}_{-k}\hat{a}_k + \Delta\hat{a}_k^*\hat{a}_{-k}^*) + \frac{|\Delta|^2}{V}, \tag{9}$$

where I have changed the integral with respect to $k$ into a discrete summation. To derive Eq. (9), I neglected terms $(\hat{a}_{-k}\hat{a}_k - \langle\hat{a}_{-k}\hat{a}_k\rangle)\cdot(\hat{a}_{k'}^*\hat{a}_{-k'}^* - \langle\hat{a}_{k'}^*\hat{a}_{-k'}^*\rangle)$, which are second order in the deviation from the mean field. Note that because $\hat{a}_k$ is independent of spin variables, the summation with respect to $k$ is lower bounded because $k \geq 0$, unlike BCS superconductor theory.

By using the relation $\xi_k = -\xi_{-k}$ ($\because \omega_k \propto k$. Cf. for fermions, $\xi_k = \xi_{-k}$), I obtain the following Hamiltonian:

$$H_{mf} = \sum_{k\geq 0}[-(\Delta\hat{a}_k^*\hat{a}_{-k}^* + \Delta^*\hat{a}_{-k}\hat{a}_k)] + \frac{|\Delta|^2}{V}, \tag{10}$$

To interpret these results, I adopt the following four postulates. (i) Plasmons are always created (annihilated) at $\omega = \mu \pm \delta\omega$ as a pair by MIs whenever they are created (annihilated); that is, the energy necessary to add one plasmon to the system is $\mu$, the chemical potential. (ii) The turbulent states are defined as the normal states with $\Delta = 0$, indicating that the phases of the LWT states are completely random. (iii) Complete dephasing (by $\pi$) of a plasmon from the phase of the mean field $\Delta$ of the other plasmon pairs is regarded as a dissociation of the plasmon pair. (iv) I use the conventional quantum interpretation of $\hat{a}_k^*$ and $\hat{a}_k$, that is, $\hat{a}_k^*$ as creation and $\hat{a}_k$ as annihilation of one plasmon, respectively. Given these postulates, the final form (RHS) of Eq. (10) can be interpreted as the change in the total energy for creation (annihilation) of a plasmon pair. The created (annihilated) plasmon pairs, whose phases are identical to the mean field $\Delta$ of the other plasmon pairs, lower the total energy $H_{mf}$ by an amount of $2|\Delta|$. The created (annihilated) plasmon pairs having antiphase $\Delta$ raise $H_{mf}$ by $2|\Delta|$. This means that the dissociation energy of a plasmon pair is $2|\Delta|$. Because $\Delta$ is proportional to the number $N_{\text{pair}}$ of plasmon pairs, the resistance to dissociation increases with increased plasmon-pair condensation. The total energy of the normal state LWT is zero because $\Delta = 0$. The energy of the ground state of the LWSC is $-|\Delta|^2/V$.

In BCS superconductor theory, the gap equation, which determines the energy gap $\Delta$ self-consistently, is obtained from Eq. (7), and the canonical transform of the Hamiltonian (Bogolubov transform) and the fact that the bogolons follow fermion statistics. In the classical LW system, another equation is needed to determine the energy gap $\Delta$ self-consistently, which is introduced by accounting for the condensation process of the plasmon pairs. To that end, I use Eq. (2). I consider an equation of motion for plasmon pairs rather than for single plasmons. By taking the product of Eq. (2) and $\hat{a}_{-k}$, I get

$$\frac{\partial \hat{a}_k}{\partial \tau}\hat{a}_{-k} = -i\xi_k \hat{a}_k \hat{a}_{-k} - iV \int \hat{a}_{-k}\hat{a}_2^* \hat{a}_3 \hat{a}_4 \delta_{k+2-3-4}\, dk_2 dk_3 dk_4. \tag{11}$$

I simplify by selecting only the pairs that satisfy the $k$-matching conditions $k + k_2 = k_3 + k_4$ and $k = -k_2, k_3 = -k_4$. This can be achieved by the replacement $\delta_{k+2-3-4} \to \delta_{k+2}\delta_{3+4} \equiv \delta(k + k_2)\delta(k_3 + k_4)$. In addition, $\hat{a}_{-k} \equiv |a_{-k}|e^{i\varphi_{-k}(t)} = |a_k|e^{-i\varphi_k(t)} = \hat{a}_k^*$ for the plasmon pairs created by the MI. I apply this replacement to $\hat{a}_{-k}\hat{a}_2^*$ on the RHS of Eq. (11) together with the above-mentioned plasmon-pair selection, resulting in $\hat{a}_{-k}\hat{a}_2^* = |\hat{a}_k|^2$. After conversion of the continuous integrations into discrete summations, $\int dk \to \sum_j 2\pi/L$, Eq. (11) becomes

$$\frac{\partial \hat{a}_k}{\partial \tau}\hat{a}_{-k} = -i\xi_k \hat{a}_k \hat{a}_{-k} - iV \int \hat{a}_{-k}\hat{a}_2^* \hat{a}_3 \hat{a}_4 \delta_{k+2}\delta_{3+4}\, dk_2 dk_3 dk_4$$

$$= -i\xi_k \hat{a}_k \hat{a}_{-k} - iV \left(\frac{2\pi}{L}\right)^3 |\hat{a}_k|^2 \sum_j (\hat{a}_j \hat{a}_{-j})e^{i(\varphi_j+\varphi_{-j})}. \tag{12}$$

Similarly, by taking the product of Eq. (2) with $-k$ and $\hat{a}_k$, I get

$$\frac{\partial \hat{a}_{-k}}{\partial \tau}\hat{a}_k = -i\xi_{-k} \hat{a}_k \hat{a}_{-k} - iV \left(\frac{2\pi}{L}\right)^3 |\hat{a}_k|^2 \sum_j (\hat{a}_j \hat{a}_{-j})e^{i(\varphi_j+\varphi_{-j})}. \tag{13}$$

By using $\xi_{-k} = -\xi_k$, the summations of Eqs. (12) and (13) lead to the phase equation of the coupled oscillators (i.e., plasmon pairs):

$$\frac{\partial \Phi_k}{\partial \tau} = -2V\left(\frac{2\pi}{L}\right)^3 \sum_j |\hat{a}_j||\hat{a}_{-j}|e^{i(\Phi_j - \Phi_k)}, \tag{14}$$

where $\Phi_k \equiv \varphi_k + \varphi_{-k}$ is the phase of the plasmon pair. For this derivation, I have assumed that $\frac{\partial |a_k|}{\partial \tau} \ll \frac{\partial \varphi_k}{\partial \tau}$. By using the distribution $g(\xi_j)$ of initial frequencies of the plasmon pairs with the normalization condition $\sum_j g(\xi_j + \xi_{-j}) = 1$, Eq. (14) may be rewritten in a so-called Kuramoto oscillator model [16]:

$$\frac{\partial \Phi_k}{\partial \tau} = (\xi_k + \xi_{-k}) + \frac{K}{N_{\text{pair}}} \sum_j \sin(\Phi_j - \Phi_k), \tag{15}$$

where $N_{\text{pair}}$ is the total number of plasmon pairs involved in the LWSC formation. I find that $\sum_j |\hat{a}_j|^2 \sim N_{\text{plasmon}} \sim 2N_{\text{pair}}$ because $|\hat{a}_j||\hat{a}_{-j}| = |\hat{a}_j|^2$, which is the number of the plasmons in mode $j$:

$$K = 4\text{Im}[V]\left(\frac{2\pi}{L}\right)^3 N_{\text{plasmon}}^2. \tag{16}$$

Note that $g(\xi_j + \xi_{-j})$ is incorporated as the initial condition in the Kuramoto model [16]. Equation (16) indicates that the interaction is enhanced as $N_{\text{plasmon}}$ increases (this is the many-body effect). The order parameter $m_{\text{Kuramoto}}$ of this system is defined in the Kuramoto model as

$$m_{\text{Kuramoto}} \equiv \frac{1}{N_{\text{pair}}} \sum_j |\hat{a}_j||\hat{a}_{-j}|e^{i\Phi_j} = \frac{\Delta}{V}. \tag{17}$$

Note that $m_{\text{Kuramoto}}$ must satisfy the self-consistent gap equation [16]:

$$m_{\text{Kuramoto}} = K m_{\text{Kuramoto}} \int_{-\pi/2}^{\pi/2} g[K m_{\text{Kuramoto}} \sin(\Phi)] \cos^2 \Phi \, d\Phi$$

$$\equiv S(m_{\text{Kuramoto}}). \tag{18}$$

This gap equation has a nontrivial solution together with a trivial solution $m_{\text{Kuramoto}} = 0$ when $K$ is larger than the critical value $K_c$. When $K < K_c$, Eq. (18) has only the trivial solution $m_{\text{Kuramoto}} = 0$, which corresponds to LWT states. $K_c$ can be obtained from $S'(0) = 1$ [16]:

$$K_c = \frac{2}{\pi g(0)}. \tag{19}$$

For a Lorentzian $g(\xi) = \gamma/\pi[(\xi - \xi_0)^2 + \gamma^2]$, $K_c = 2\gamma$.

To verify the Kuramoto oscillator condensation of the plasmon pairs, I measure the order parameter $m_{\text{Kuramoto}} = \Delta/V$ in our PIC simulations. The order parameter $m_{\text{Kuramoto}}$ is calculated as follows, which is equivalent to Eqs. (7), (8), and (17):

$$m_{\text{Kuramoto}} = \frac{\langle \sum_{\omega_1=\mu_F}^{\mu_F+\omega_D}[E_x(\omega_1)E_x(2\mu_F-\omega_1)]\rangle}{\langle \sum_{\omega_1=\mu_F}^{\mu_F+\omega_D}\sqrt{|E_x(\omega_1)|^2|E_x(2\mu_F-\omega_1)|^2}\rangle}. \quad (20)$$

The order parameter $m_{\text{Kuramoto}}$ is the fraction of the plasmon pairs whose center-of-mass wavenumber and phases are $k_F$ and identical to those of the mean field $\Delta$. Thus, $m_{\text{Kuramoto}}$ lies in the range $0 \leqq |m_{\text{Kuramoto}}| \leqq 1$. When all the plasmon pairs have the same phase, $m_{\text{Kuramoto}}$ is unity. Figure 3(c) clearly shows the existence of a critical value in $N_{\text{plasmon}} \propto \sqrt{K}$, which agrees with the above theory. This theory can explain as well the discontinuous transition of the maximum tricoherence of the surface elevation of the capillary rogue waves in laboratory from the weak phase coupling in four-wave interactions to the strong one when the forcing of the water container exceeded the threshold [7].

In the proposed theory, temperature of the system is not defined because although the whole system including the waves and the particles (ions and electrons) is in an equilibrium state, the subsystems are apparently not in thermal equilibrium. For example, the electron distribution functions in the LWSC state are far from the Maxwell distribution, as shown in Fig. 5. A plateau is observed in the high-velocity range, which corresponds to the FWM band. Here the relationship $v = \omega/k$ is used. It is clear that the LWSC is already Landau damped because $df/dv = 0$ in the FWM band. Thus, the LWSC is, in terms of plasma physics, a so-called Bernstein–Greene–Kruskal mode, which is characterized by undamped large-amplitude waves [17].

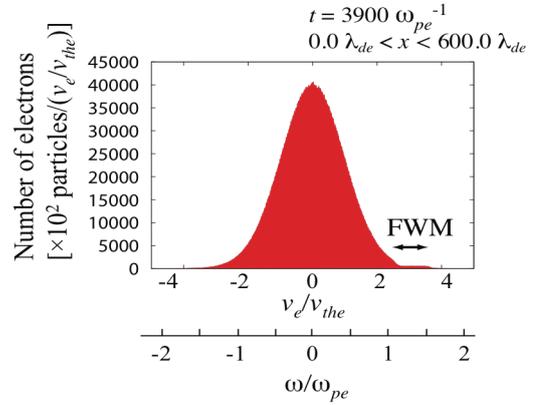

Fig. 5. Spatially averaged velocity distribution function of electrons for LWSC state. The FWM-frequency band is indicated. The relationship $v = \omega/k$ is used. A plateau forms at the FWM band, indicating that the LWSC is a phase-mixed Bernstein–Greene–Kruskal mode.

In summary, a transition between turbulent and coherent supercontinuum (SC) states of classical LWs observed in a 1D PIC simulation is explained by a classical version of the BCS superconductor theory. Supercontinuum and turbulent states both exhibit broad power spectra, whereas only SC states have significant spatiotemporal coherence. The MI and resultant creation of plasmon pairs are the classical counterparts of the Cooper instability and the creation of Cooper pairs in superconducting states. Bose–Einstein condensation of Cooper pairs in superconducting states is replaced by Kuramoto oscillator-entrainment of plasmon pairs in LWSC states. As the total number of plasmons increases, LW systems transit from a turbulent state to a coherent SC state. The total number of plasmons is proportional to the strength of the interaction between plasmons. This transition is accompanied by an increase in the order parameter, which clearly indicates that a macroscopic number of plasmon pairs occupy a single momentum state with an identical phase in the LWSC. In addition to the order parameter, a classical version of the von Neumann entropy and the spatiotemporal coherence of the waves show a clear transition between the two phases when the total wave quanta of the plasmons increase, indicating spontaneous symmetry breaking and the emergence of long-range order. In terms of plasma physics, the LWSC is a so-called Bernstein–Greene–Kruskal mode, which is characterized by undamped large-amplitude waves. The discovery of the first classical counterpart of the superconductor transition indicates existence of quantum-classical correspondence of the BCS superconductor phase transition.

## ACKNOWLEDGMENTS

This work was supported by Grants-in-Aid MOST 104-2112-M-006-014-MY3 from the Ministry of Science and Technology, Taiwan.